\documentclass[aps,preprint]{revtex4}
\usepackage{amsfonts}
\usepackage{amsmath}
\usepackage{amssymb}
\usepackage{graphicx}

\input{tcilatex}

\begin{document}

\title{Asymmetrical two-atom entanglement in a coated microsphere}
\author{G.Burlak$^{1}$, A.Klimov$^{2}$}
\affiliation{$^{1}$Center for Research on Engineering and Applied Sciences, Autonomous
State University of Morelos, Cuernavaca, Mor. 62210, Mexico.
gburlak@uaem.mx, $^{2}$Departamento de Fisica, Universidad de Guadalajara,
Revolucion 1500, Guadalajara, Jalisco, 44420, Mexico. klimov@cencar.udg.mx}

\begin{abstract}
We study evolution of entanglement of two two-level atoms placed inside a
multilayered microsphere. We show that due to inhomogeneity of the field
modes this entanglement essentially depends on the atomic positions
(asymmetrical entanglement) and also on the detuning between the atomic
transitions and field frequencies. The robust and complete entanglement can
be achieved even in the resonant case when the atoms have different
effective coupling constants, and it can be extended in time if the detuning
is large enough. We study analytically the lossless case and estimate
numerically the effect of dissipative processes.
\end{abstract}

\maketitle

\section{Introduction}

Recently, essential progress in fabrication and determination of optical
properties of different kinds of microcavities with sizes about $0.1-20\mu m$
which contain semiconductor nanoclusters or quantum dots (QDs) has been
achieved [see \cite{TakashiYamasaki:2000a}, \cite{Artemyev:2000a} and
references therein]. When semiconductor QDs are embedded in a spherical
microcavity, the QD luminescence can be coupled with eigen modes of the
electromagnetic field of the microcavity and a lower threshold of stimulated
emission (or lasing modes) of QDs can be achieved. In recent papers \cite%
{Artemyev:2000a},\cite{Artemyev:2001a},\cite{RuiJia:2001a} a coupling
between the optical emission of embedded $CdSe_{x}S_{1-x}$ , QDs and
spherical cavity modes was studied and a strong whispering gallery mode
(WGM) resonance with high $Q$ factors is registered in the photoluminescence
spectra. Recently \cite{Moller:2003a} quantum-confined semiconductor
nanorods were used as highly polarized nanoemitters for active control of
the polarization state of microcavity photons.

Until now the modes with small numbers of spherical harmonics (SNM) are
essentially less well-studied compared with whispering gallery modes (WGM)
due to their rather low $Q$-factor caused by significant radiating losses. A
possibility of strongly increasing the $Q$-factor of the microsphere was
proposed in several papers, see e.g. \cite{DavidBrady:1993a}, \cite%
{Sullivan:1994a}, \cite{Burlak:2000a}. The main idea consists in coating a
microsphere by alternative layers of a spherical stack, which results in an
increase of the $Q$ factor up to values comparable for WGM, i.e. $%
10^{7}-10^{9}$.

In a system with small mean photon number two spatially separated atoms in a
cavity become entangled at some time moments as a result of sharing the
re-radiated photons \cite{Phoenix:1993a}, \cite{Kudryavtsev:1993a}, \cite%
{Cirac:1994a}, \cite{Freyberger:1996a}, \cite{Plenio:1999a}. One of these
schemes has been realized using Rydberg atoms coupled one by one to a high $%
Q $ microwave superconducting microcavity \cite{Hagley:1997a}. In
inhomogeneous structures, like multilayered microspheres, the quantized
field properties are quite different from the unbounded case because of a
non-uniformity of the cavity field, which becomes important for the
entanglement dynamics. In spite of numerous experimental obstacles, mainly
related with the decoherence problem, it seems very natural to entangle
atoms placed in high-$Q$ cavities (like microspheres) via interaction with
modes of the cavity quantized field. A simple scheme for the generation of
two-atom maximally entangled states via dispersive interaction was proposed
in \cite{Shi-BiaoZheng:2000a}. A number of papers report studies of the
evolution of the entanglement in an atomic subsystem resonantly interacting
with a single mode of the cavity field (two-atom Tavis-Cummings model) \cite%
{Tessier:2003a}, \cite{Shi-BiaoZheng:2005a}, \cite{Shang-BinLi:2005a}. A
robust generation of many-particle entanglement in various configurations
has been discussed in \cite{Unanyan:2002a}, \cite{AliCan:2003a}, \cite%
{Ficek:2002a}. Experimentally a robust entanglement was recently studied in%
\cite{Haffner:2005a} where an entanglement lasting for more than $20s$ was
observed in a system of two trapped $Ca+$ ions. Authors \cite%
{Gao-xiangLi:2004b} have shown that the degree of entanglement between the
two atoms strongly depends on the mean photon number and the strength of
two-photon correlations.

In \cite{Dung:2002a} a scheme for entangling $N$ two-level atoms located
close to the surface of a dielectric microsphere and atoms resonantly
interacting with the field was considered. It was shown that in the
particular case of two atoms located at diametrically opposite positions a
perfect entanglement cannot be achieved even in the strong-coupling regime.

In this paper we study two-atom entanglement interacting with field modes
inside a microsphere covered with spherical dielectric alternating layers
(coated microsphere). We are mainly interested: \textit{(i)} in the
frequency range of the high reflectivity field in $\lambda/4$ -stack; 
\textit{(ii)} in the case when identical atoms are located asymmetrically
inside the microsphere (i.e. the system is not symmetric with respect to a
permutation of initially excited and non-excited atoms), so that the field
inhomogeneity leads to different effective atom-field coupling constants ;
and \textit{(iii)} the atomic transitions can be both resonant and well
detuned from the field peak frequency.

The paper is organized as follows. In Section II we discuss basic equations
for two atoms placed into a coated microsphere and the solution for this
case. In Section III we present an analytical solution for probability
amplitudes and apply it to studying the atomic concurrence. In Section IV we
present a numerical study of the concurrence (tangle) dynamics. In the last
Section, we discuss and summarize our conclusions.

\section{Basic equations}

Consider two identical two-level atoms coupled to a quantized
electromagnetic cavity field in a coated microsphere (Fig.\ref%
{Pic_greenfiggeom1b}).

\FRAME{ftbpFU}{6.0156in}{4.0845in}{0pt}{\Qcb{Geometry of coated microsphere
with two atoms.}}{\Qlb{Pic_greenfiggeom1b}}{greenfiggeom1b.eps}{\special{
language "Scientific Word"; type "GRAPHIC"; maintain-aspect-ratio TRUE;
display "USEDEF"; valid_file "F"; width 6.0156in; height 4.0845in; depth
0pt; original-width 10.6666in; original-height 7.2307in; cropleft "0";
croptop "1"; cropright "1"; cropbottom "0"; filename
'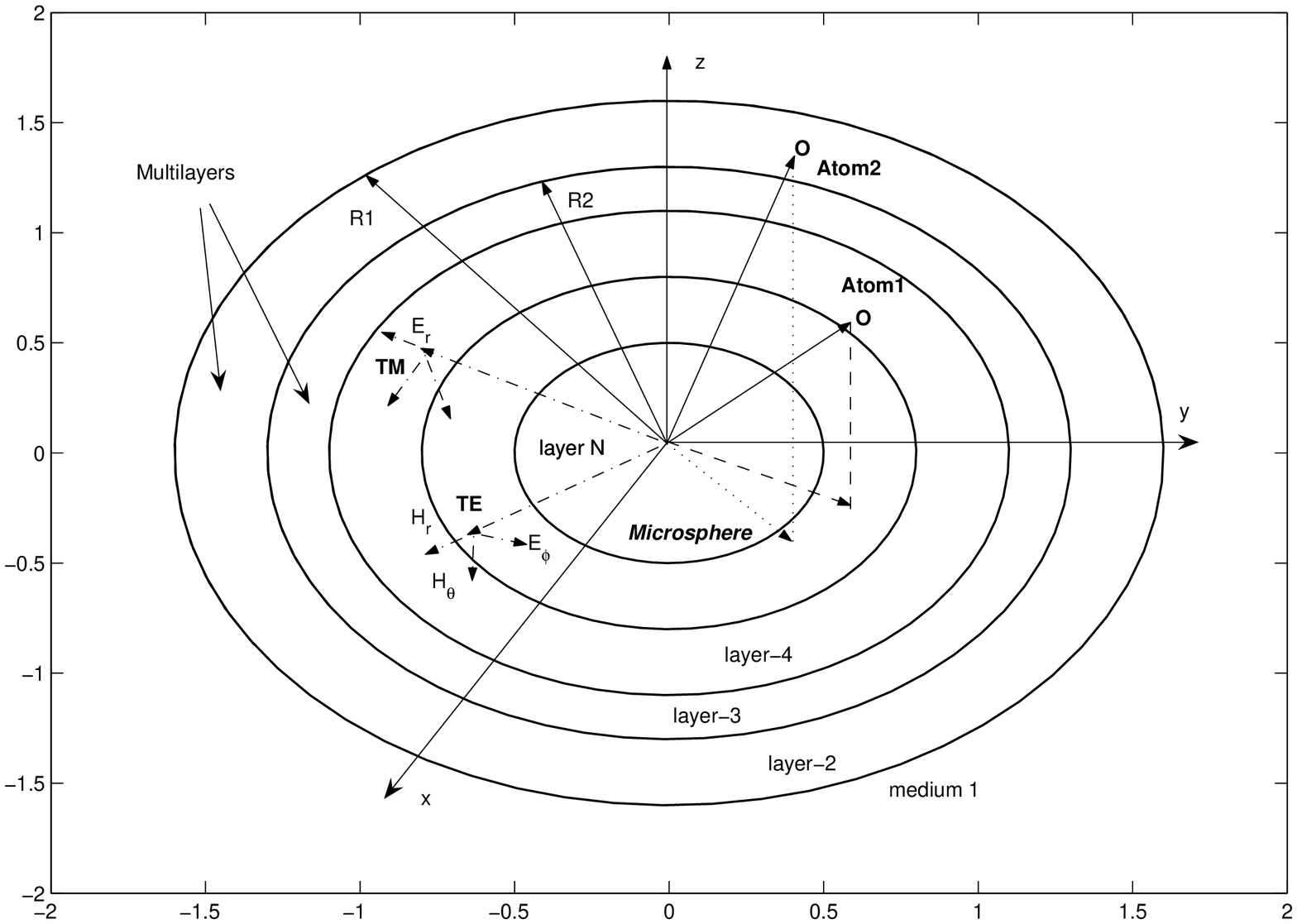';file-properties "XNPEU";}}

Let us assume that the atoms are sufficiently far from each other, so that
the interatomic Coulomb interaction can be ignored. In this case, the
electric dipole and rotating wave approximations can be applied and the
Hamiltonian for the atom-cavity system ($\hbar=1$) is given by \cite%
{HoTrungDung:2000a}, \cite{Gruner:1996a}

\begin{equation}
H=H_{0}+H_{1}\text{,}  \label{General H}
\end{equation}%
\begin{align*}
H_{0} & =\hat{H}=\int\!\mathrm{d}^{3}{\mathbf{r}}\!\int_{0}^{\infty
}\!\!d\omega\,\hbar\omega\,\hat{f}^{\dagger}({\mathbf{r}},\omega){}\hat {f}({%
\mathbf{r}},\omega)+\sum_{j=1,2}{{\frac{1}{2}}}\omega_{j}\widehat{s}_{jz}%
\text{, } \\
H_{1} & =-\sum_{j}[\widehat{s}_{j}^{\dagger}\hat{E}^{(+)}({\mathbf{r}}_{j}){}%
{d}_{j}+H.c.]\text{, }
\end{align*}
where $\omega_{j}$\ is the atomic transition frequency ($\omega_{1}=\omega
_{2}=\omega_{at}$) , $s_{z,\pm j}$, $j=1,2$ are the atomic operators
corresponding to the $j$-th atom and obeying standard $su(2)$ commutation
relations, $\left[ s_{\pm},s_{z}\right] =\pm s_{\pm},\left[ s_{+},s_{-}%
\right] =2s_{z}$, $d_{j}$ are atomic dipoles. Here $\hat{f}(r,\omega)$ and $%
\hat{f}^{\dagger}(r,\omega)$ are bosonic operators which play the role of
the fundamental variables of the electromagnetic field and the medium,
including a reservoir necessarily associated with losses in the medium. The
electric-field operator is expressed in terms of $\hat{f}(r,\omega)$ as \cite%
{HoTrungDung:2000a}, \cite{Gruner:1996a}, 
\begin{equation}
\hat{E}^{(+)}(\mathbf{r})=i\sqrt{\frac{\hbar}{\pi\varepsilon_{0}}}\int
_{0}^{\infty}d\omega\frac{\omega^{2}}{c^{2}}\int\mathrm{d}^{3}\mathbf{r}%
^{\prime}\sqrt{\varepsilon_{\mathrm{I}}(\mathbf{r}^{\prime},\omega)}\mathbf{G%
}(\mathbf{r},\mathbf{r}^{\prime},\omega){}\hat{f}(\mathbf{r}^{\prime
},\omega),  \label{El.field}
\end{equation}
with 
\begin{align}
\left[ \hat{f}_{i}({\mathbf{r}},\omega),\hat{f}_{j}^{\dagger}({\mathbf{r}}%
^{\prime},\omega^{\prime})\right] & =\delta_{ij}\delta(\mathbf{r}-\mathbf{r}%
^{\prime})\delta(\omega-\omega^{\prime})\text{, }  \label{ff comm} \\
\left[ \hat{f}_{i}(\mathbf{r},\omega),\hat{f}_{j}(\mathbf{r}%
^{\prime},\omega^{\prime})\right] & =0=\left[ \hat{f}_{i}^{\dagger}({\mathbf{%
r}},\omega),\hat{f}_{j}^{\dagger}({\mathbf{r}}^{\prime},\omega^{\prime})%
\right] \text{,}  \notag
\end{align}
where $G(\ r,\ r^{\prime},\omega)$ is the classical Green tensor satisfying
the equation 
\begin{equation}
\left[ \frac{\omega^{2}}{c^{2}}\,\epsilon({\ \mathbf{r}},\omega )-\mathbf{%
\nabla}\times\mathbf{\nabla}\times\right] \mathbf{G}({\ \mathbf{r}},{\ 
\mathbf{r}}^{\prime},\omega)=-\delta({\ \mathbf{r}}-{\ \mathbf{r}}^{\prime})
\label{GreenEq}
\end{equation}
together with the boundary condition at infinity [$\delta(\mathbf{r})$ is
the dyadic $\delta$-function].\ Here $\varepsilon(\mathbf{r}%
,\omega)=\varepsilon _{\mathrm{R}}(\mathbf{r},\omega)+i\varepsilon_{\mathrm{I%
}}(\mathbf{r}^{\prime },\omega)$\ is the complex dielectric permittivity. We
look for the solution of the Schr\"{o}dinger equation with the Hamiltonian (%
\ref{General H}) in a single excitation manifold in the form 
\begin{align}
\lefteqn{|\Psi(t)\rangle=C_{1}(t)\left\vert 0\right\rangle \left\vert
e_{1}g_{2}\right\rangle +C_{2}(t)\left\vert 0\right\rangle \left\vert
g_{1}e_{2}\right\rangle +}  \label{Psi full} \\
& +|g_{1}g_{2}\rangle\int\mathrm{d}^{3}\mathbf{r}\int_{0}^{\infty}d\omega%
\,[C_{3i}(\mathbf{r},\omega,t)|\{1_{i}(\mathbf{r},\omega)\}\rangle ]\text{,}
\notag
\end{align}
where $|e_{k}\rangle$ ($|g_{k}\rangle$) denotes the excited (ground) atomic
state of $k$-th atom. Correspondingly, $|\{1_{i}(r,\omega)\}\rangle=\!\hat {f%
}_{i}^{\dagger}(r,\omega)|\{0\}\rangle$ is a single photon Fock state and, $%
|\{0\}\rangle$ is the vacuum state of the rest of the system. Note that this
state is not a photonic state in general, but a state of the macroscopic
medium dressed by the electromagnetic field \cite{LandauStatistical:1981a},%
\cite{Wylie:1984a},\cite{Gruner:1996a},\cite{Dung:2002a}.

\FRAME{ftbpFU}{4.6951in}{3.5362in}{0pt}{\Qcb{(a), (b) and (c). Frequency
spectrum of imaginary parts of tangential component of the dyadic Green's
function $\func{Im}(G_{\protect\varphi \protect\varphi }(r,r^{\prime },f))$, 
$f=\protect\omega /2\protect\pi $ for $7$-layered system (microsphere coated
with $5$ alternating $\protect\lambda /4$\ layers), with atomic positions $%
a_{1}=0.9\protect\mu m$ and $a_{2}=1.1\protect\mu m$. Refraction indexes of
the layers are $n_{4}=1.5+i2\cdot 10^{-4}$ (glass, bottom microsphere, $1%
\protect\mu m$), $n_{3}=3.58+i10^{-3}$($Si$, $0.12\protect\mu m$), $%
n_{2}=1.46+i3\cdot 10^{-3}$ ($SiO_{2}$, $0.3\protect\mu m$) and $n_{1}=1$
(surrounding space). (a) $\func{Im}(G_{\protect\varphi \protect\varphi %
}(a_{1},a_{1},f))$; (b) $\func{Im}(G_{\protect\varphi \protect\varphi %
}(a_{1},a_{2},f))$; and (c) $\func{Im}(G_{\protect\varphi \protect\varphi %
}(a_{2},a_{2},f))$; (d) radial dependence of $\func{Im}(G_{\protect\varphi 
\protect\varphi }(r,a_{2},f_{f}))$, where the atom is placed in $a_{2}=0.9%
\protect\mu m$ and the field's peak frequency is $f_{f}\ \ =241.7THz$. Dash
line in (d) shows the refraction indexes of the spherical stack structure.}}{%
\Qlb{Pic_fig2_fr}}{fig2_fr.eps}{\special{language "Scientific Word";type
"GRAPHIC";maintain-aspect-ratio TRUE;display "USEDEF";valid_file "F";width
4.6951in;height 3.5362in;depth 0pt;original-width 5.834in;original-height
4.3863in;cropleft "0";croptop "1";cropright "1";cropbottom "0";filename
'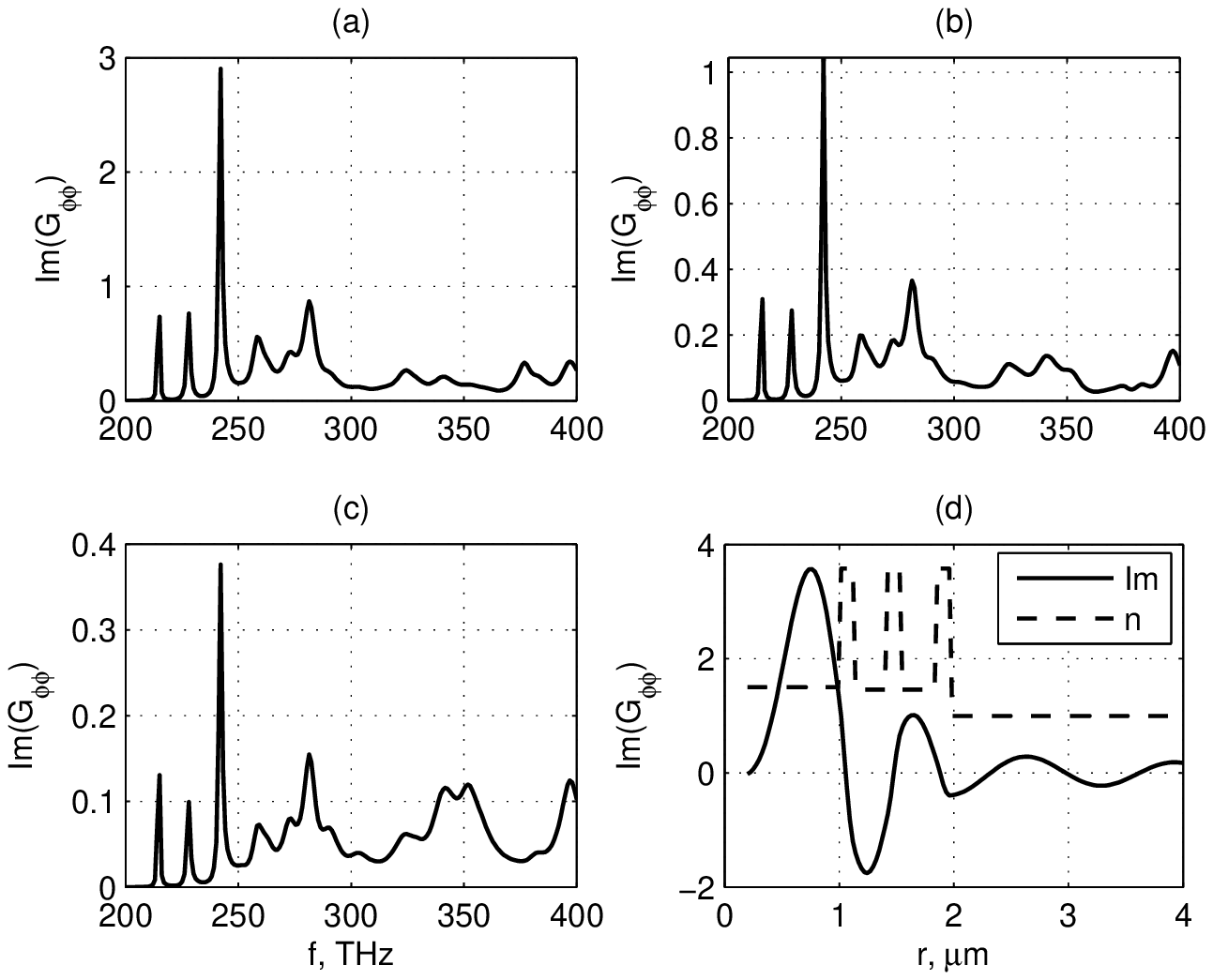';file-properties "XNPEU";}}For simplicity we study the
frequency range close to the microsphere resonance with the frequency $%
\omega _{f}$, when the Green function can be written as 
\begin{equation}
G(\mathbf{r},\mathbf{r}^{\prime },\omega )=G(\mathbf{r},\mathbf{r}^{\prime
},\omega )\cdot \delta (\omega -\omega _{f}).  \label{Green delta}
\end{equation}

The effect of broadening of such a line due to dissipation is studied in
Sec.IV numerically. For further references we show in Fig.\ref{Pic_fig2_fr}
the typical frequency spectrum of Green's function, calculated numerically.
Let us assume that the atomic dipoles are parallel to the surface of the
microsphere (similar to the situation considered in \cite{Moller:2003a}), so
only tangential components of Green's tensor (e.g. $\mathbf{G}\varphi\varphi$%
) give a contribution.

Projecting $|\Psi(t)\rangle$ in (\ref{Psi full})\ onto $\left\vert
0\right\rangle \left\vert e_{i}g_{k}\right\rangle $\ and $%
|\{1_{i}(r,\omega)\}\rangle\left\vert g_{1}g_{2}\right\rangle $ states, we
obtain the following equations for the probability amplitudes $C_{i}$:

\begin{equation}
\dot{C}_{1,2}(t)=-iB(a_{1,2},\omega_{f},t)\text{, }
\label{Eqs for C1,2 and B}
\end{equation}

\begin{equation*}
\dot{B}(\mathbf{r},\omega_{f},t)=i\Delta\omega B(\mathbf{r},\omega _{f},t)-i%
\overline{\mathbf{G}}(\mathbf{r},a_{1},\omega_{f})\text{ }C_{1}(t)-i%
\overline{\mathbf{G}}(\mathbf{r},a_{2},\omega_{f})\text{ }C_{2}(t)\text{,}
\end{equation*}
where $\mathbf{r}$ is coordinate vector, $a_{1,2}$ are the positions of the
atoms in the microsphere, $\Delta\omega=\omega_{f}-\omega_{at}$, and

\begin{align}
B(\mathbf{r},\omega_{f},t) & =d_{i}\int\mathrm{d}^{3}\mathbf{r}^{\prime }{%
\cdot}\alpha\,\mathbf{G}_{ik}(a,\mathbf{r}^{\prime},\omega_{f})C_{3k}(%
\mathbf{r}^{\prime},\omega_{f},t)\text{,}  \label{B} \\
\overline{\mathbf{G}}(\mathbf{a},\mathbf{r}^{\prime},\omega) & =\kappa
d_{i}d_{k}\func{Im}(\mathbf{G}_{ik}(\mathbf{a},\mathbf{r}^{\prime },\omega))%
\text{,}
\end{align}
where $\alpha=i\sqrt{\varepsilon_{I}(\mathbf{r},\omega_{f})/\pi\varepsilon
_{0}}\omega_{f}^{2}/c^{2}$ and $\kappa=\omega_{f}^{2}/c^{2}\pi%
\varepsilon_{0} $. Eliminating $B(\mathbf{r},\omega_{f},t)$\ from (\ref{Eqs
for C1,2 and B}) we obtain after minor algebra closed equations for $%
C_{1,2}(t)$ in matrix form as follows%
\begin{equation}
\frac{d^{2}\mathbf{q}}{dt^{2}}\mathbf{-}i\Delta\omega\frac{d\mathbf{q}}{dt}+%
\mathbf{A\cdot q}=0\text{,}  \label{Eqs for C1,2 Matr}
\end{equation}
where

\begin{equation*}
\mathbf{q=}%
\begin{bmatrix}
C_{1} \\ 
C_{2}%
\end{bmatrix}
\text{, \ }\mathbf{A=}%
\begin{bmatrix}
\overline{G}(1,1) & \overline{G}(1,2) \\ 
\overline{G}(2,1) & \overline{G}(2,2)%
\end{bmatrix}
\text{.}
\end{equation*}
To derive (\ref{Eqs for C1,2 Matr}) the identity \cite{HoTrungDung:2000a} $%
\func{Im}\,G_{kl}(\mathbf{r,r\prime},\omega)=\int\mathrm{d}^{3}\mathbf{s}%
\left( \omega^{2}/c^{2}\right) \varepsilon_{\mathrm{I}}(\mathbf{s}%
,\omega)G_{km}(\mathbf{r},\mathbf{s},\omega)G_{lm}^{\ast }(\mathbf{r}%
^{\prime},\mathbf{s},\omega)$ was taken into account. From now on we adopt
the convention of summation over repeated vector-component indices.

The general solution of Eq.(\ref{Eqs for C1,2 Matr}) has the form%
\begin{equation}
C_{k}(t)=\sum_{j=1}^{4}c_{kj}e^{i\omega_{j}t}\text{, }k=1,2\text{,}
\label{GenrSolCi}
\end{equation}
where the frequencies $\omega_{j}$ are solutions of the eigenvalue problem $%
\det\left[ \left( -\omega^{2}+\omega\Delta\omega\right) \delta_{kl}+A_{kl}%
\right] =0$ or

\begin{equation}
\left( -\omega^{2}+\omega\Delta\omega\right) ^{2}+\left(
-\omega^{2}+\omega\Delta\omega\right) Tr\{\mathbf{A}\}+\det(\mathbf{A})=0%
\text{.}  \label{eigValProbl}
\end{equation}
Following \cite{Le-WeiLi:1994a} we rewrite the Green tensor for a
multilayered microsphere as follows 
\begin{equation}
\mathbf{G}(\mathbf{r},\mathbf{r^{\prime}},\omega)=\mathbf{G}^{V}(\mathbf{r},%
\mathbf{r}^{\prime},\omega)\delta_{fs}+\mathbf{G}^{(fs)}(\mathbf{r},\mathbf{r%
}^{\prime},\omega),  \label{GreenFull}
\end{equation}
where $G^{\mathrm{V}}(r,r^{\prime},\omega)$ represents the contribution of
the direct waves from the radiation sources in an unbounded medium, $f$ and $%
s$ denote the layers where the field point and source point are located, $%
\delta_{fs}$ is the Kronecker symbol, and the scattering Green tensor $%
\mathbf{G}^{(fs)}(\mathbf{r},\mathbf{r}^{\prime},\omega)$ describes the
contribution of both multiple reflection and transmission. The Green tensor $%
\mathbf{G}^{(fs)}$\ in general can be expanded as

\begin{equation}
\mathbf{G}^{(fe)}(\mathbf{r},\mathbf{r}^{\prime},\omega)=\frac{ik_{s}}{4\pi }%
\sum_{p=e,o}\sum_{n=1}^{\infty}\sum_{m=0}^{n}\frac{2n\!+\!1}{n(n\!+\!1)}%
\frac{(n\!-\!m)!}{(n\!+\!m)!}(2\!-\!\delta_{0m})\mathbf{G}_{pnm}^{\left(
f,e\right) }(\mathbf{r},\mathbf{r}^{\prime},\omega)\text{,}
\label{Green refl}
\end{equation}
where $G_{cnm}^{\left( f,e\right) }(\mathbf{r},\mathbf{r\prime},\omega)$ is
a particular Green tensor, $n$ is the spherical and $m$ is the azimuth
quantum numbers of a microsphere, $k_{i}=\omega n_{i}/c$, $n_{i}=\sqrt{%
\varepsilon _{i}(\omega)}\,$\ is a refraction index.$\,\ $General recurrent
formulas and particular representations of the Green tensor $G_{pnm}^{\left(
f,e\right) }(\mathbf{r},\mathbf{r\prime},\omega)$ can be found in Ref.\cite%
{Le-WeiLi:1994a}.

Generally analysis of the Green tensor (\ref{Green refl}) requires intensive
computation. In the simplest case when atoms are located at positions with
the same value of the field amplitude, we have $\overline{G}(i,j)=\overline{G%
}$ and $\det(\mathbf{A})=0$. This case is symmetrical with respect to
permutation of the atoms and one can easily obtain the solution $%
\omega_{1}=0 $, $\omega_{2}=\Delta\omega$ and $\omega_{3,4}=\left(
\Delta\omega/2\right) \pm\left[ \left( \Delta\omega/2\right) ^{2}+2\overline{%
G}\right] ^{1/2}$. However, experimentally, a symmetric location of the
atoms with respect to the center of the microsphere is difficult to achieve 
\cite{Plenio:1999a}. In the simplest nontrivial case we have to take into
account the nonuniformity of the field. In this case we have $\overline{G}%
(1,1)\neq\overline{G}(2,2)$\ , but the coefficients $\overline{G}(i,k)$ can
be written as follows

\begin{equation}
\overline{G}(1,1)=\chi_{1}^{2}\text{, \ }\overline{G}(2,2)=\chi_{2}^{2}\text{%
, }\overline{G}(1,2)=\overline{G}(2,1)=\chi_{1}\chi_{2}\text{,}
\label{G and g}
\end{equation}
so that the condition

\begin{equation}
\overline{G}(1,1)\cdot\overline{G}(2,2)=\overline{G}(1,2)^{2}  \label{G*G=GG}
\end{equation}
is fulfilled. We have again $\det(\mathbf{A})=0$, but now $\omega
_{3,4}=\left( \Delta\omega/2\right) \pm\Omega$, $\Omega^{2}=\left(
\Delta\omega/2\right) ^{2}+\overline{G}(1,1)+\overline{G}(2,2)$, so that the
atoms have different Rabi frequencies. Introducing the new variable $C_{3}$
according to

\begin{equation}
\dot{C}_{3}-i\Delta\omega C_{3}=-i\left( \chi_{1}C_{1}+\chi_{2}C_{2}\right) 
\text{,}  \label{eq.for C3}
\end{equation}
the system (\ref{Eqs for C1,2 Matr}) can be reduced to a simple form

\begin{equation}
\dot{C}_{1}=-i\chi_{1}C_{3}\text{, }\dot{C}_{2}=-i\chi_{2}C_{3}.
\label{simpleSystC12}
\end{equation}
From direct calculations we have found that such a case is fulfilled for a
spherical structure with a $7$-layered system (microsphere coated with $5$
alternating $\lambda/4$\ layers), see Fig.\ref{Pic_fig2_fr} with $%
R_{2}=1.97\mu m$, $R_{1}=1\mu m$. Two atoms having tangentially oriented
dipoles $\mathbf{d}\perp$ $\widehat{\mathbf{r}}$ are at positions $%
a_{1}=0.9\mu m$ and $a_{2}=1.1\mu m$ correspondingly. In Fig.\ref%
{Pic_fig2_fr}(d) we show the radial distribution of the Green tensor
component $\func{Im}(\mathbf{G}_{\varphi\varphi})$ for the first atom in the
microsphere. We have found that for this structure $\chi_{1}=1.72$, $%
\chi_{2}=0.608$ and $\chi_{2}/\chi_{1}=0.35$. As $\chi_{1}>$ $\chi_{2}$ we
can see that in this case the coupling constant is larger for the first atom
or, in other words, the first atom is placed in a stronger field mode of the
coated microsphere.

\section{Effective Hamiltonian dynamics}

In the lossless case the Hamiltonian corresponding to the simplified
situation in the single-mode regime described by Eq.(\ref{Green delta}) can
be represented in the following form

\begin{equation}
H=\omega_{f}a^{\dagger}a+\omega_{at}\left( s_{z1}+s_{z2}\right) +\chi
_{1}\left( as_{+1}+h.c.\right) +\chi_{2}\left( as_{+2}+h.c.\right) \text{,}
\label{Simple H}
\end{equation}
where the effective coupling constants $\chi_{i}=\overline{G}(i,i)^{1/2}$
depend on the positions of atoms inside the microsphere. Because the
coupling constant is larger for the atom placed in the region of a stronger
field mode, the configuration is not symmetrical with respect to the
permutation of the atoms. In this case the state vector is given by

\begin{equation}
|\Psi(t)\rangle=C_{1}(t)\left\vert 0\right\rangle \left\vert
e_{1}g_{2}\right\rangle +C_{2}(t)\left\vert 0\right\rangle \left\vert
g_{1}e_{2}\right\rangle +C_{3}(t)\left\vert 1\right\rangle \left\vert
g_{1}g_{2}\right\rangle \text{,}  \label{Simple Psi}
\end{equation}
where $C_{1,2}(t)$ are solutions of the Eqs.(\ref{eq.for C3}),(\ref%
{simpleSystC12}):

\begin{equation}
C_{1}(t)=-\chi_{1}r(t)+\lambda\text{, }C_{2}(t)=-\chi_{2}r(t)+1-\lambda\text{%
, }C_{3}(t)=-i\left( \chi_{\lambda}/\Omega\right) \exp(i\Delta\omega
t/2)\sin(\Omega t)\text{,}  \label{SimpleSolC123}
\end{equation}
with%
\begin{align}
r(t) & =\left( \chi_{\lambda}/\chi^{2}\right) \{\exp(i\Delta\omega t/2)\left[
i\frac{\Delta\omega}{2\Omega}\sin(\Omega t)-\cos(\Omega t)\right] +1\}\text{,%
}  \label{r(t)} \\
\chi_{\lambda} & =\chi_{1}\lambda+\chi_{2}(1-\lambda)\text{, }%
\chi^{2}=\chi_{1}^{2}+\chi_{2}^{2}\text{,}  \notag
\end{align}
and the initial conditions $C_{1}(0)=\lambda$, $C_{2}(0)=1-\lambda$, ($%
\lambda=1,0$) are considered (for the $\lambda=0$ case the system evolves
from the initial state $\left\vert g_{1}e_{2}\right\rangle $) . In
particular, the average photon number can be easily calculated using the
solution (\ref{SimpleSolC123}):%
\begin{equation*}
\left\langle n\right\rangle =\left\vert C_{3}(t)\right\vert ^{2}=\left(
\chi_{\lambda}/\Omega\right) ^{2}\sin^{2}(\Omega t).
\end{equation*}
The reduced atomic density matrix for the state (\ref{Simple Psi}) has the
form

\begin{equation}
\rho ^{a}=Tr_{f}\{\left\vert \Psi \right\rangle \left\langle \Psi
\right\vert \}=%
\begin{bmatrix}
0 & 0 & 0 & 0 \\ 
0 & \left\vert C_{1}\right\vert ^{2} & \left\vert C_{1}C_{2}\right\vert & 0
\\ 
0 & \left\vert C_{1}C_{2}\right\vert & \left\vert C_{2}\right\vert ^{2} & 0
\\ 
0 & 0 & 0 & \left\vert C_{3}\right\vert ^{2}%
\end{bmatrix}%
.  \label{ReducMatrDens1}
\end{equation}%
In the frame of the standard approach \cite{Wootters:1998a} we obtain from (%
\ref{ReducMatrDens1}) the concurrence $C(t)$ for two atom system as%
\begin{equation}
C(t)=2\left\vert C_{1}C_{2}\right\vert \text{.}  \label{Conc gener}
\end{equation}%
In Fig.\ref{Pic_fig1_c2} \ we show the dynamics of tangle $C^{2}$\ (see Eq.(%
\ref{Conc gener})) for the cases $\lambda =0$ (Fig.\ref{Pic_fig1_c2}(a)),
and $\lambda =1$ (Fig.\ref{Pic_fig1_c2}(b)). For this configuration the
condition $\chi _{1}>\chi _{2}$ is fulfilled, in Fig.\ref{Pic_fig1_c2}(a)
the first atom being in the ground state is placed in a larger field
strength. In this case the tangle $C^{2}$\ has\ the form of well resolved
periodical plateau. The inverse situation is shown in Fig.\ref{Pic_fig1_c2}%
(b). In Fig.\ref{Pic_fig1_c2} the dashed line shows the average photon
number $\left\langle n\right\rangle $. One can observe that the amplitude of
rapid oscillations in the upper part of a long periodical tangle evolution
(plateau) $C^{2}$ is essentially less in the case when an unexcited atom is
placed in the region of stronger field (Fig.\ref{Pic_fig1_c2}(a)). It is
easy to see, that such oscillations have the Rabi frequency and are related
to the instantaneous average number \ of photons (dashed line in Fig.\ref%
{Pic_fig1_c2}) stored in the field.

\FRAME{ftbpFU}{4.6855in}{3.5241in}{0pt}{\Qcb{Dynamics of the two-atom tangle 
$C^{2}$ vs $\protect\tau =\protect\omega _{at}t$ for $\protect\chi %
_{1}=0.254 $ and $\protect\chi _{2}=0.151$ in a lossless case for (a)
initial state $\left\vert g_{1}e_{2}\right\rangle $; (b) initial state $%
\left\vert e_{1}g_{2}\right\rangle $. Due to large detuning $\Delta \protect%
\omega /\protect\omega _{at}\simeq 0.5$ the amplitude of mean photon number $%
\left\langle n\right\rangle $ oscillations is less in (a) case. One can see
the well-recognized plateaus of tangle in case (a). In case (b) the plateaus
is adding with strong oscillations due to large $\left\langle n\right\rangle 
$.}}{\Qlb{Pic_fig1_c2}}{fig1c2.eps}{\special{language "Scientific Word";type
"GRAPHIC";maintain-aspect-ratio TRUE;display "USEDEF";valid_file "F";width
4.6855in;height 3.5241in;depth 0pt;original-width 5.8219in;original-height
4.3708in;cropleft "0";croptop "1";cropright "1";cropbottom "0";filename
'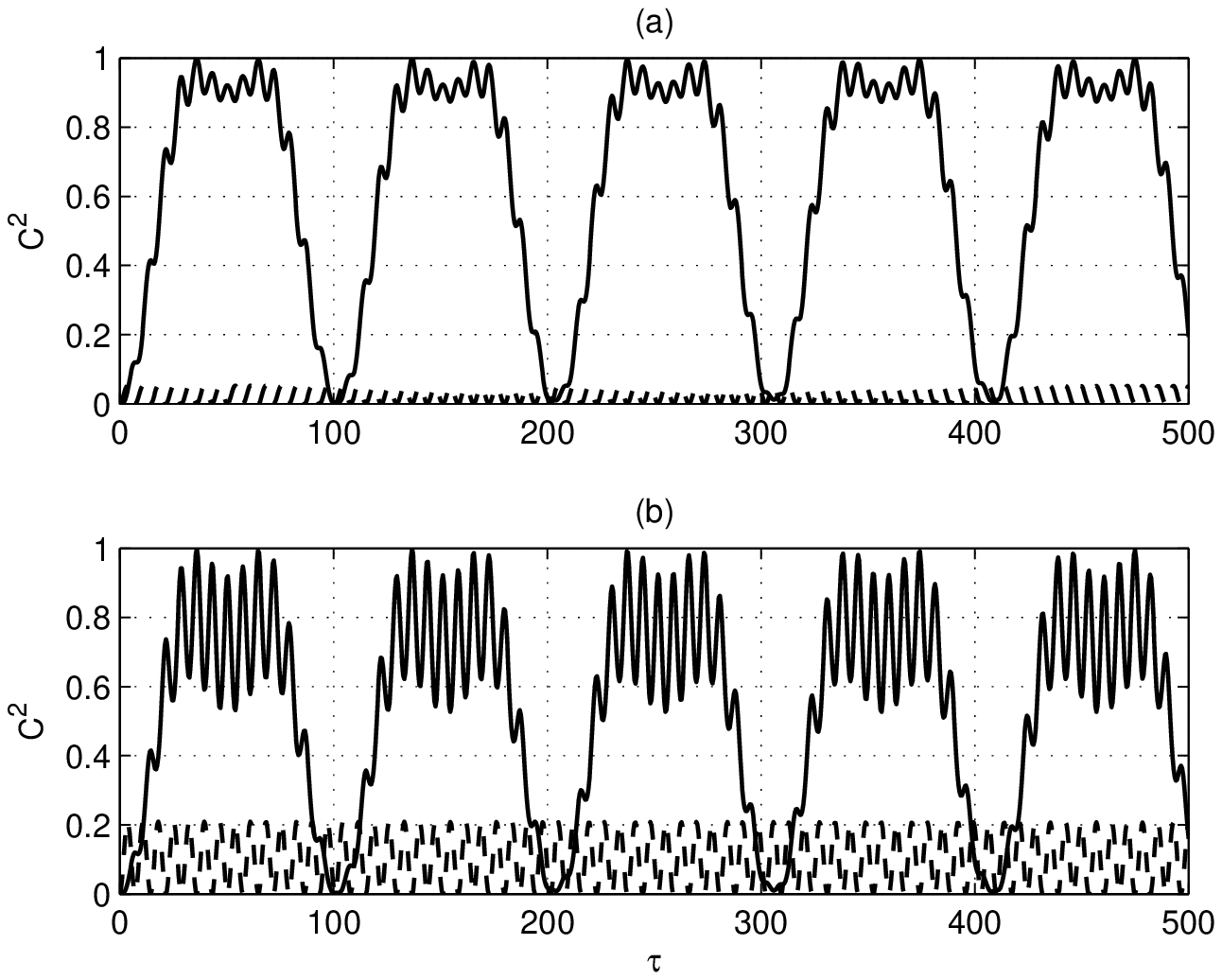';file-properties "XNPEU";}}

\FRAME{ftbpFU}{4.6855in}{3.5241in}{0pt}{\Qcb{Tangle $C^{2}$\ vs $\protect%
\tau =\protect\omega _{at}t$ in general case for parameters $\Delta \protect%
\omega /\protect\omega _{at}=0.75$ , and (a) $G(1,1)=0.01$, $%
G(1,2)=G(2,1)=0.012$ and $G(2,2)=0.04$. Dash line shows tangle for $%
G(1,1)=0.01$, $G(1,2)=G(2,1)=0.02$ and $G(2,2)=0.04$ when Eq.(\protect\ref%
{G*G=GG}) is valid. In case (b) $G(1,1)\leftrightarrows G(2,2)$.}}{\Qlb{%
pic_c01_04_075}}{c01_04_075.eps}{\special{language "Scientific Word";type
"GRAPHIC";maintain-aspect-ratio TRUE;display "USEDEF";valid_file "F";width
4.6855in;height 3.5241in;depth 0pt;original-width 5.8219in;original-height
4.3708in;cropleft "0";croptop "1";cropright "1";cropbottom "0";filename
'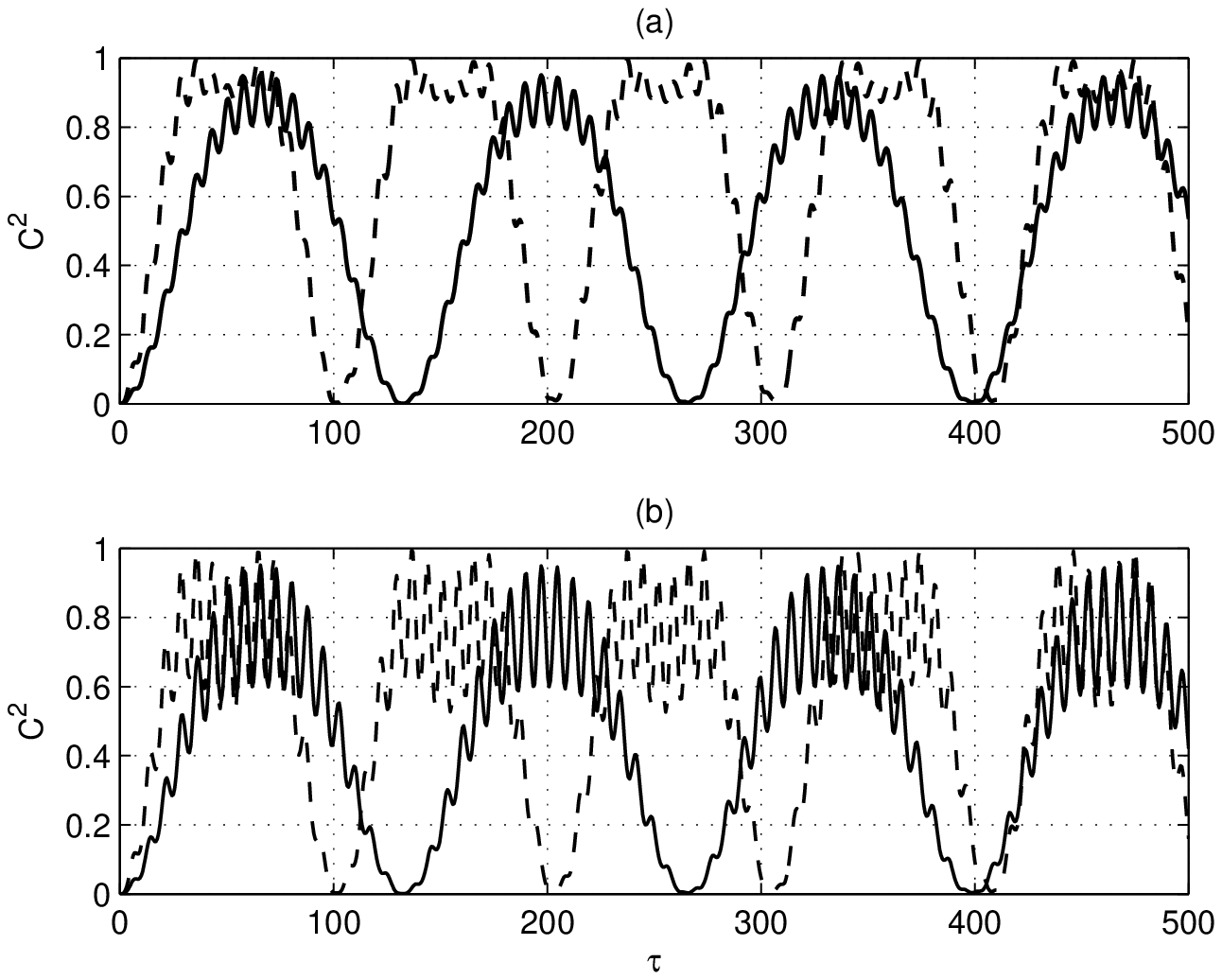';file-properties "XNPEU";}}

It is worth noting that in the case when \ $\overline{G}(1,1)\cdot\overline {%
G}(2,2)\neq\overline{G}(1,2)^{2}$ the general formulae (\ref{Eqs for C1,2
Matr}) should be used to study the dynamics of concurrence $C$. The
evolution of the concurrence for the general situation is shown in Fig.\ref%
{pic_c01_04_075} (solid line). The particular situation described by the
simplified model (\ref{SimpleSolC123}) is presented in the same Fig.\ref%
{pic_c01_04_075} as the dashed line. From Fig.\ref{pic_c01_04_075} one can
observe that if (\ref{G*G=GG}) is fulfilled the amplitude of the fast
oscillations is much less than in the general case, and therefore can be
regarded as the optimal dynamics.

For the far detuned case $|\Delta\omega|>>\chi$ one can easily obtain from (%
\ref{SimpleSolC123}) two well separated frequency components of $C_{1,2}(t)$%
:the high frequency component $\exp(i\Delta\omega t)$ and the low frequency
component $\exp(ig^{2}t/\Delta\omega)$. The latter leads to a formation of
well recognized plateaus, which do not exist in the resonant case. Further,
we will use the initial conditions corresponding to $\lambda =1$, so that $%
\chi_{\lambda}=\chi_{1}$. The concurrence (\ref{Conc gener}) for such a
solution can be easily analyzed in parameter space $\chi_{1},\chi_{2}$\ for
the case $\Delta\omega=0$. In this case concurrence (\ref{Conc gener}) is
explicitly asymmetrical with respect to the atomic permutation ($\chi
_{1}\rightleftarrows\chi_{2}$) and has the form

\begin{equation}
C(\chi_{1},\chi_{2})=2\frac{\chi_{1}\chi_{2}}{\Omega^{2}}\left\vert 1-\frac{%
\chi_{1}^{2}}{\Omega^{2}}\left[ 1-\cos(\Omega t)\right] \right\vert \cdot%
\left[ 1-\cos(\Omega t)\right] .  \label{Concur g1g2}
\end{equation}
It is easy to see from (\ref{Concur g1g2}) that the surface $%
C(\chi_{1},\chi_{2})$ is separated by circles (with radii $\left(
\chi_{1}^{2}+\chi _{2}^{2}\right) ^{1/2}=2k\pi/t$, $k=0,1,2..$) on which $%
C(\chi_{1},\chi _{2})=0$ (atoms are disentangled). The detailed structure of
the concurrence can be better understood rewriting Eq.(\ref{Concur g1g2}) in
the form

\begin{equation}
C(k,a)=2\frac{ak}{1+k^{2}}\left\vert 1-\frac{a}{1+k^{2}}\right\vert \text{,}
\label{C(k,a)}
\end{equation}%
where $k=\chi _{2}/\chi _{1}$ and the quantity $a=1-\cos (\Omega t)$ is in
the range $0\leq a\leq 2$. At a fixed value of $a$ the concurrence $C(k,a)$
assumes maximal values $C(k,a)\leq 1$\ at $k_{1,2}=2^{-1}\left[ 6a\pm
2\left( 9a^{2}-4a+4\right) ^{1/2}\right] ^{1/2}$, and $C(k,a)=1$ only at $%
a=2 $ when $k^{\pm }=\sqrt{2}\pm 1=1/k^{\mp }$. Note that $k_{2}$ exists for 
$a\geq 1$. In general $1\leq k_{1}\leq k^{+}\approx 2.41$ and $0\leq
k_{2}\leq k^{-}\approx 0.41$. This means that in the resonant case, $\Delta
\omega =0$, the two-atom system can be maximally entangled if $\chi
_{2}/\chi _{1}=k^{\pm }$ \ , i.e. when the atoms have different field-atom
coupling constants, $\chi _{2}\neq \chi _{1}$. Nevertheless in the general
case, when $\Delta \omega \neq 0$\ , the structure of the concurrence $%
C(\chi _{1},\chi _{2})$ is more complicated.

The structure $C(\chi _{1},\chi _{2})$ was calculated for $\omega _{at}t=27$%
, $\lambda =0$ (initially excited atom is placed in a smaller field) and $%
\Delta \omega /\omega _{at}=0.5$. We observe that the surface $C(\chi
_{1},\chi _{2})$ is rather asymmetrical with respect to the line $\chi
_{1}=\chi _{2}$. In the course of evolution for fixed $\chi _{1},\chi _{2}$
the maximal values of $C(\chi _{1},\chi _{2})$ move out from the origin of
coordinates. Obviously on the edges where $\chi _{1},\chi _{2}=0$, the
concurrence vanishes, $C(\chi _{1},\chi _{2})=0$. In the vicinity of maxima
the concurrence $C(\chi _{1},\chi _{2})$ is highly asymmetric. For $\chi
_{1}>\chi _{2}$ the maxima $C(\chi _{1},\chi _{2})$ in the left side are
smoother and the hills are more pronounced. This means that the system
remains in the region of strong entanglement for long periods if $\chi
_{1}>\chi _{2}$. However in general, the details of the surface $C(\chi
_{1},\chi _{2})$ essentially depend on $\Delta \omega $ and the form of the
Green's function.

\section{Numerical study}

It is worth noting, that in a real microsphere the field dissipation is
caused by material losses and the radiation into surrounding space leads to
line broadening (bandwidth). The analytical calculation of such a broadening
requires an extensive knowledge of the microscopical local field, which in a
multilayered microsphere case is itself a quite difficult problem. To
estimate the influence of the dissipation on the concurrence dynamics we
will use the following simplified approach. Although we do not know the
exact frequency dependence of the dissipative part on the refractive indices
of the materials $n_{i}$ in a microsphere, it is possible to calculate the
spectral width of the Green function peak.\emph{\ }Thus, we can estimate the
effect of the field's dissipation using the master equation technique in the
framework of the Lingblad approach. In particular, the dissipation
coefficients are calculated from the bandwidth of the Green function peak%
\emph{\ }(see Fig.\ref{Pic_fig2_fr}). Such a semi-analytical approach allows
us to simulate numerically not only the evolution of the concurrence in a
lossy environment, but also the dynamics of the average photon number. In
this approach we replace the exact Hamiltonian (\ref{General H}) by the
simplified Hamiltonian (\ref{Simple H}) and numerically solve the following
master equation for the joint atom-field density operator $\rho $ in a
dissipative cavity at zero temperature:

\FRAME{ftbpFU}{4.6855in}{3.5241in}{0pt}{\Qcb{The same as in Fig.\protect\ref%
{Pic_fig1_c2} but for loss case ($\protect\gamma =2\cdot 10^{-2}$). To see
the details of long-time dynamics we calculate $C^{2}$ up to $\protect\tau %
_{\max }=1000$.}}{\Qlb{Pic_fig1_c2diss}}{fig2c2diss.eps}{\special{language
"Scientific Word";type "GRAPHIC";maintain-aspect-ratio TRUE;display
"USEDEF";valid_file "F";width 4.6855in;height 3.5241in;depth
0pt;original-width 5.8219in;original-height 4.3708in;cropleft "0";croptop
"1";cropright "1";cropbottom "0";filename '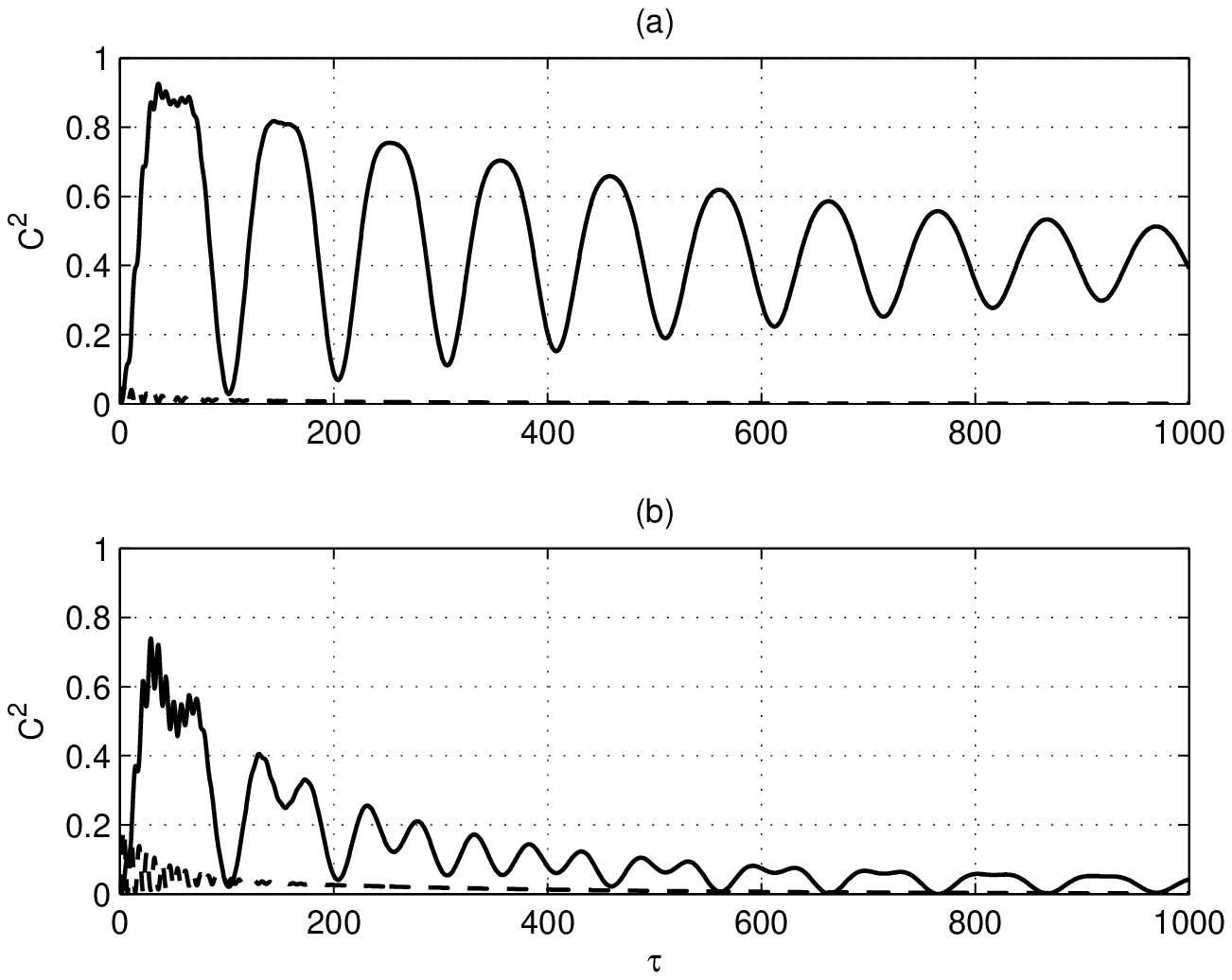';file-properties
"XNPEU";}}

\begin{align}
\frac{d\rho}{d\tau} & =-i[H,\rho]+L_{1}\rho\text{,}  \label{Neiman eq} \\
L_{1}\rho & =\gamma_{1}\left( 2a\rho a^{+}-a^{+}a\rho-\rho a^{+}a\right) 
\text{,}
\end{align}
where $H$ is given by (\ref{Simple H}), and in (\ref{Neiman eq}) we neglect
the atomic spontaneous emission in the Rabi period time scale. Also we have
used the detuning $\Delta\omega/\omega_{at}=0.5$, and $\chi_{1}=0.254$, $%
\chi_{2}=0.151$, $\chi_{2}/\chi_{1}=0.594$. In Fig.\ref{Pic_fig1_c2diss} the
dynamics of the two-atom tangle $C^{2}$ for the lossy case is shown. It is
clear from Fig.\ref{Pic_fig1_c2diss} that the plateaus of concurrence
survive even in the presence of dissipation, although their amplitude is
obviously lower than in the lossless case.

\section{Conclusion}

In conclusion, we have studied the dynamics of entanglement of spatially
separated two-level atoms interacting with a radially nonuniform cavity
field mode in a dielectric microsphere coated with an alternating stack. We
found that due to the field inhomogeneity the atoms can be maximally
entangled even in the resonant case. We have found that entanglement
essentially depends on the atomic positions (asymmetrical entanglement) and
also on the detuning between atoms and the field mode frequencies. The
entanglement is considerably more stable with duration much longer than the
period of Rabi oscillations (robust entanglement) when the unexcited atom is
placed in a stronger field, while the detuning increases the duration of the
entanglement period. The dissipation reduces the amplitude of the
entanglement, however practically does not change the width of the zones of
large entanglement.

\section{Acknowledgements}

This work of G.B is partially supported by CONACyT grant 47220. The work of
A.K. is partially supported by CONACyT grant 45704.

\end{document}